\documentclass[rnote]{aa} 
\usepackage{txfonts}
\usepackage{graphicx}
\usepackage{natbib}

\bibpunct{(}{)}{;}{a}{}{,}

\newcommand{\oam}{\mbox{\object{AM\,Her}}}
\newcommand{\ohz}{\mbox{\object{HZ43}}}
\newcommand{\amher}{AM\,Her}

\newcommand{\rosat}{\textit{ROSAT}}
\newcommand{\chandra}{\textit{Chandra}}

\begin{document}

\title{Soft X-ray in-flight calibration of the \rosat\ PSPC\thanks{Based on
observations made with the \rosat\ satellite.}}

\author {K.\ Beuermann \inst{1}}


\institute{Institut f\"ur Astrophysik, Friedrich-Hund-Platz 1,
D-37077 G\"ottingen, Germany, beuermann@astro.physik.uni-goettingen.de}

\date{Received January 5, 2008 / accepted February 11, 2008}
 
\authorrunning{K. Beuermann et al.}  \titlerunning{Soft X-ray spectral
variability of AM Herculis}

\abstract {}
{ 
We present an in-flight calibration of the \rosat\ PSPC using the
incident spectra of the hot white dwarf HZ43 and the polar \amher.}
{ 
We derive an absolute flux calibration of the PSPC using the accurately
known soft X-ray spectrum of HZ43.  Corrections to the PSPC response matrix
are derived from a comparison of predicted and observed PSPC spectra
of HZ43, supplemented by results for \amher.}
{ 
The calibration of the PSPC for photon energies $E<0.28$\,keV is found
to be accurate to better than 5\% refuting earlier reports of a major
miscalibration. Our corrections to the detector response matrices
remove systematic residuals in the pulse height spectra of soft
sources.  }
{}
\keywords {Methods: data analysis -- Stars: individual: (\ohz) -- Stars: individual: (\oam) -- X-rays:stars}

\maketitle

\begin{figure*}[t]
\includegraphics[width=10.49cm]{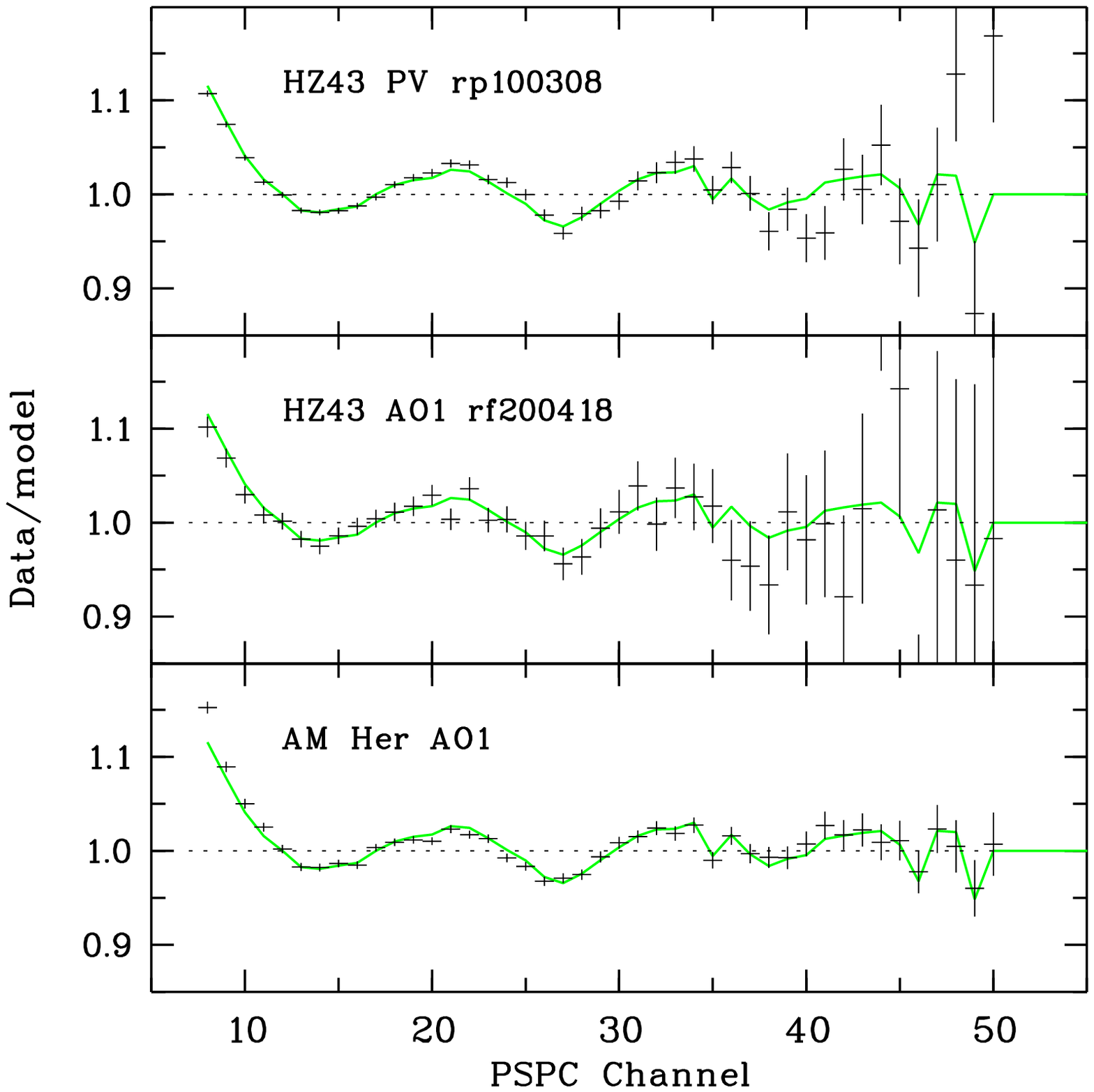}
\hspace{6mm}
\includegraphics[width=6.521cm]{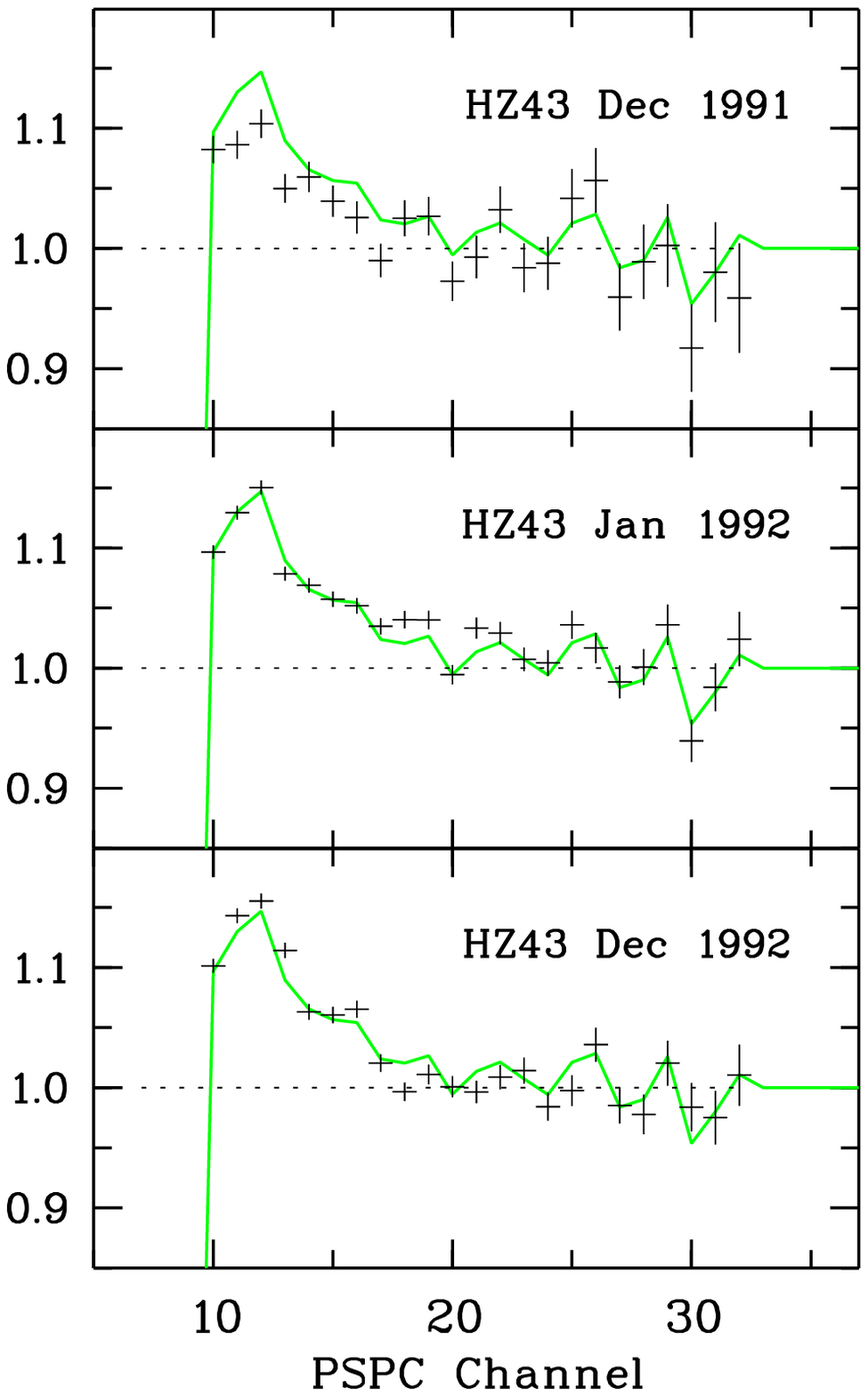}
\caption[ ]{Residuals of the PSPC spectra calculated for the incident
spectra of the calibrator star HZ43 and the polar \amher, derived from
the \chandra\ LETG+HRC spectra with the LETG effective area correction
of \citet{beuermannetal06} included. One PSPC channel corresponds to a
nominal apparent photon energy of 10\,eV. The green curves represent
the weighted averages of the residuals for the PSPC observations
before 11 October 1991 (left panel) and after that date (right
panel). They are used to create the corrected matrices DRMPSPC-AO1c and
DRMPSPCc.}
\label{fig:drm}
\end{figure*}

\section{Introduction}

The absolute flux calibration of satellite instruments in the soft
X-ray regime is complicated by the lack of suitable flux
standards. The situation is better in the keV range and much better in
the optical and ultraviolet regimes, where the observed and model
spectral fluxes of white dwarfs agree at the 1\% level
\citep[e.g.][]{bohlinetal01}. Instruments with spectral capability in
the soft X-ray regime are usually calibrated in flight by comparing
the observed spectra of white dwarfs with pure hydrogen atmospheres
with theoretical spectra extrapolated from optical and ultraviolet
wavelengths. The absolute flux calibration of the \rosat\ PSPC differs
in that it is based entirely on ground calibrations
\citep[e.g.][]{brieletal97,schmittsnowden90}. Several authors have
claimed that observations of white dwarfs indicated an actual
sensitivity of the PSPC that is lower than nominal by a factor of
about two \citep{napiwotzkietal93,jordanetal94,wolffetal96,wolffetal99}.
A decisive step was the establishment of HZ43 and Sirius\,B as soft
X-ray standards by \citet{beuermannetal06}. They fitted model spectra
for the two white dwarfs and the neutron star RX\,J1856-37\footnote{A
two-blackbody model for RXJ\,1856-27 and TMAP spectra for HZ43 and
Sirius\,B. For the latter see
http://astro.uni-tuebingen.de/\raisebox{.2em}{\small
$\sim$}rauch/TMAP/UserGuide/ UserGuide.html} simultaneously to the
\chandra\ LETG+HRC spectra of the three stars obtaining the incident
spectra along with a common wavelength-dependent correction to the
effective area of LETG+HRC spectrometer. Based on these results, I
present an absolute flux calibration of the PSPC and a correction to
the detector response.

\section{Absolute calibration of the \rosat\  PSPC}

\begin{table*}[t]
\caption{Journal of \rosat\ PSPC observations of HZ43 and \amher\ used
in the present paper, supplemented by the fit values for $g$, $w$, and
$b$ (see text). A downward arrow denotes that a parameter is kept at
its previous value. The observed count rates are integrals for channel
11 to 60.}
\label{tab:obslog}
\begin{tabular}{l@{\hspace{3mm}}lcrcccccc} 
\hline \hline \noalign{\smallskip}
Object &  Identifier & Date &  Exp (s) & Counts/s & PSPC & open/boron & $g$ & $w$ & $b$ \\           
\noalign{\smallskip} \hline
\noalign{\smallskip}
\ohz & rp100308      &900621 & 21520 & $66.09\pm0.09$  & C & \hspace{-1.8mm}open & $0.986\pm0.003$ & $0.988\pm0.015$ & --\\
     & rf100113      &900621 &   290 & \hspace{1.8mm}$8.65\pm0.27$ & C & boron & $\downarrow $ & $\downarrow $ & $0.943\pm0.025$\\
     & rp200418      &910619 &   262 & $64.62\pm0.77$  & B & \hspace{-1.8mm}open & $0.993\pm0.003$ & $0.963\pm0.015$ & \\
     & rf200418      &910619 & 21625 & \hspace{1.8mm}$8.04\pm0.03$ & B & boron & $\downarrow $ & $\downarrow $ & $0.990\pm0.015$\\
     & rp141808,09,12&911211 &  1411 & $62.23\pm0.32$  & B & \hspace{-1.8mm}open & $0.998\pm0.003$ & $\downarrow $ &  \\
     & rp141818,27,28&920112 &  5420 & $63.73\pm0.16$  & B & \hspace{-1.8mm}open  & $0.998\pm0.005$ & $\downarrow $ &  \\
     & rf141822      &920113 &  1424 & \hspace{1.8mm}$8.12\pm0.12$ & B & boron & $1.005\pm0.005$ & $\downarrow $ & $\downarrow $ \\
     & rp141916,7    &921215 &  6901 & $62.68\pm0.17$  & B & \hspace{-1.8mm}open  & $0.988\pm0.003$ & $\downarrow $ & \\
     & rf141917      &921215 &  2796 & \hspace{1.8mm}$7.83\pm0.08$ & B & boron & $\downarrow$ & $\downarrow $ & $\downarrow $ \\
     & rf141500      &930709 &  2663 & \hspace{1.8mm}$7.68\pm0.08$ & B & boron & $0.978\pm0.003$ & $\downarrow $ & $\downarrow $ \\
AM\,Her& rp300067-0  &910412 & 9394&\hspace{-1.8mm}$134.27\pm0.13$ & B & \hspace{-1.8mm}open  & $0.990\pm0.003$ & $\downarrow $ & \\
\noalign{\smallskip} \hline      
			         
\end{tabular}
\end{table*}

Folding the incident spectrum of HZ43 through the PSPC response with
the nominal effective area and detector response matrix\footnote{The
nominal detector response matrices, \mbox{DRMPSPC-AO1} and DRMPSPC,
the effective areas of both PSPCs, and the correction vectors drived
here, $c_\mathrm{AO1}(x)$ and $c(x)$, are available via
http://www.mpe.mpg.de/xray/wave/rosat/doc/calibration/index.php. We
use the nominal effective areas of PSPC~C and PSPC~B as given by
effarea-pspcc and effarea-pspcb except for a minor correction that
more accurately represents the structure of the carbon K-edge at
43.6\AA\ (M. Freyberg, private communication).} yields PSPC spectra
that closely resemble the observed ones \citep[][see their
Fig.~9]{beuermannetal06}.

The remaining differences in flux (count rate) and spectral shape
disappear if we choose values of the gain $g$ of the PSPC and the
thicknesses (wavelength-dependent optical depths) $w$ of the detector
window and $b$ of the boron filter that differ from nominal. Nonlinear
systematic gain variations over the lifetime of the PSPC were
discussed by \citep{prietoetal96}. At the low energies considered here
(channels $x=8-60$), it suffices to consider a linear gain change that
compresses or stretches the response matrix $D(x,E)$ for a given
incident energy $E$ by an energy-independent, but possibly
time-dependent, factor $g$. We consider $w$ or $b$ as time-independent
but, of course, they may differ for PSPC C and B used before and after
25 January 1991, respectively. \mbox{DRMPSPC-AO1} is the nominal
detector response matrix for PSPC~C and PSPC~B for observations taken
before 11 October 1991, when the high voltage of PSPC~B was reduced,
and DRMPSPC is recommended after that date.

For the observations with both PSPCs before 11 October 1991, nearly
perfect agreement between the predicted and the observed PSPC spectra
of HZ43 is obtained for values of the gain $g$, the window thickness
$w$, and the boron filter thickness $b$ a few percent less than
nominal (Table~1, lines 1 to 4). The errors quoted for these
quantities include the effect of anti-correlated variations in $w$ and
$b$ and the 5\% uncertainty in the absolute flux of HZ43~A
\citep[][Table.~3]{beuermannetal06}. We conclude that the absolute
calibration of the PSPC is very close to nominal. In fact, both PSPCs
may be slightly more sensitive than nominal, as indicated by
$w,b<1$. Keeping $w$ and $b$ fixed for observations with PSPC~B after
11 October 1991, we obtain good fits for a relative gain that
decreases slightly with time, in agreement with the results of
\citet{prietoetal96}. A relative gain $g\la1$ is expected from the
reduced sensitivity at the center of the PSPC \citep{snowdenetal01} as
the standard location for pointed observations. The central gain
depression was present from the begining of the mission, but varies
with time. It has a depth of about 4\%, but the effect is mitigated by
the $\pm3$\,arcmin systematic oscillation of the pointing direction
that moves the source back and forth across the depression with a
400\,s period.

\section{Corrections to the detector response matrix}

The low energy resolution of the PSPC should produce smooth spectra,
whereas the observed ones still contain residuals that become obvious
if the statistical errors are sufficiently small. The residuals
between predicted and observed spectra are shown in Fig.~\ref{fig:drm}
(left panel) for the HZ43 observations before 11 October 1991.  The
same undulations with an amplitude of a few percent occur in both
observations taken with PSPC~B and PSPC~C. We also show the residuals
for AM Herculis computed with the \chandra\ LETGS observation No. 6561
with the corrected effective areas of the LETGS
\citep{beuermannetal06} and applying an energy-independent
normalization factor of 1.03 to account for a small difference in the
relative flux levels of \amher\ in the \chandra\ and the PSPC
observations. That the latter factor is so close to unity is
fortuitous considering the variable character of the source. Within
statistical errors, \amher\ shows the same residuals as HZ43. The
green curve underlying the data points for the three observations is
their weighted mean. The green curve defines a correction vector
$c_\mathrm{AO1}(x)$ for $x\le50$, with $c_\mathrm{AO1}(x)\equiv 1.0$
for $x>50$ and $x$ the channel number. To correct a PSPC spectrum
taken prior to 11 October 1991, we either divide the observed spectrum
by $c_\mathrm{AO1}(x)$ or multiply all rows (constant incident photon
energy) of the detector response matrix DRMPSPC-AO1 with
$c_\mathrm{AO1}(x)$ to obtain the corrected matrix DRMPSPC-AO1c. The
observed residuals may result from non-linearities in the
analog-to-digital converter and possibly from remaining uncertainties
in the pulse height distributions for low-energy photons that have
their centroid below the detector threshold. The residuals may,
therefore, not be strictly independent of photon energy and we advise
to use DRMPSPC-AO1c preferably for sources as soft as white
dwarfs. This is the case for supersoft X-ray sources and for most
polars observed with \rosat.

A correction of this type has already been applied to the detector
response matrix DRMPSPC valid for all observations after 11 October
1991 \citep{brieletal97}. Not surprisingly, therefore, the undulations
present in the residuals of the early HZ43 observations are gone in
exposures after that date (Fig.~\ref{fig:drm}, right panel). The only
remaining significant feature is a pile-up of counts at channel
numbers below 20 peaking near threshold, which resembles the final
rise below channel 13 in the left panel. We apply this second-order
correction vector $c(x)$ to the detector response matrix DRMPSPC valid
after 11 October 1991 to create the corrected matrix DRMPSPCc. The
caveat concerning harder sources applies here, too. In principle, a
possible energy dependence of the features in both panels of
Fig.~\ref{fig:drm} could be studied using sources with different, but
known, incident spectra. The only source we are aware of, however, that
is reasonably constant has sufficient exposure in the PSPC, and a
well-exposed \chandra\ LETG spectrum is Capella.

\begin{acknowledgements} We thank Vadim Burwitz for giving us the 
mean \chandra\ LETG spectrum of \amher\ and Michael Freyberg for
supplying the gain maps of the \rosat\ PSPC. We thank Ulrich Briel,
Frank Haberl, J\"urgen H.~M.~M. Schmitt, Joachim Tr\"umper, and other
former and present colleagues at the MPE Garching for enlightening
discussions on the properties of the \rosat\ PSPC and its calibration.

\end{acknowledgements}

\bibliographystyle{aa}

\end{document}